\documentclass{elsart}
\usepackage[final]{epsfig}
\usepackage{amsmath}
\usepackage{graphics}

\newcommand{\be}{\begin{equation}}
\newcommand{\ee}{\end{equation}}
\newcommand{\beas}{\begin{eqnarray*}}
\newcommand{\eeas}{\end{eqnarray*}}
\newcommand{\bea}{\begin{eqnarray}}
\newcommand{\eea}{\end{eqnarray}}
\newcommand{\req}[1]{(\ref{#1})}

\begin{document} 

\begin{frontmatter}

\title{Aggregating partial, local evaluations to achieve global ranking}

\author {Paolo Laureti, Lionel Moret, Yi-Cheng Zhang}
\address{Department of Physics, University of Fribourg, Chemin du Musée 3, CH - 1700 Fribourg, Switzerland}

\begin{abstract}

We analyze some voting models mimicking online evaluation systems
intended to reduce the information overload.
The minimum number of operations needed for a system
to be effective is analytically estimated.
When herding effects are present, linear preferential attachment
marks a transition between trustful and biased reputations.
\end{abstract}


\end{frontmatter}

\section{The General Problem}
Electronic communities of all kinds, transaction systems in particular,
bring together users otherwise unknown to each other. In order for expressed
opinions to be trusted or for transactions to materialize,
rating systems are often crucial to rely upon \cite{MaZh04,LaSlYu02}. 
They collect information about participants' past
behavior, aggregate them and display the result. 
Such systems become most effective if
posted ratings are weighted according to raters' reputation, the prototype
application of this feedback technique being Epinions. 

Previous work \cite{ChSi01,FuNi03,ZaMoMa99}
concentrates on optimal reputation
systems, assuming that users pick the object to be judged
from a uniform distribution
and that the process converges to the real values, without
bothering about the amount of resources spent. In
reality we have to face a limited amount of evaluations
and biased probability distributions of the number of votes.

Here we shall consider a system of $N$ objects endowed with an intrinsic
quality $q_i, \; i=1,2,...,N$. At each time step $t$, a user evaluates
(or ranks) $k$ objects drawn from a probability distribution $p(t)$.
We shall analyze different situations of increasing complexity
within this framework, 
trying to find out under what circumstances the original values $\vec{q}=[q_1,q_2,...,q_N]$
(or their ordering) can be reconstructed from the aggregation of 
the evaluations with non zero probability for $N \to \infty$. 

\section{Random evaluations}
\label{secRE}
In this section we consider the case where objects to be evaluated
are chosen from a uniform distribution $p_i(t)=1/N$.

{\bf Occupancy problem.}
As a first example we fix $k=1$ and ask what is the minimum number
of users $t_1$ needed to evaluate every object at least once with
a given probability $P>0$.
The probability that after $t$ steps all objects
have been evaluated reads  \cite{Fe68v1}:
$$
P(t,N)= \sum_{j=0}^N (-1)^j \binom{N}{j} \left(1-\frac{j}{N} \right)^t 
\underset{t,N \to \infty}{\sim}  e^{- \lambda},
$$
where $\lambda \simeq N e^{-\frac{t}{N}}$ is the expected number 
of unjudged objects. The above limit becomes greater
than zero as soon as the number of evaluators $t$ reaches
the value
\be\label{t1}
t_1 \simeq N \log \frac{N}{\lambda}.
\ee
%
§
%

{\bf Pairwise comparisons.}
Let us now fix $k=2$ and assume that players can just
compare (not rate)
one randomly chosen couple at each
step. We call $r_i$ the $i^{th}$ ranked agent according
to her intrinsic value $q_i$.
If raters make no mistakes, the real rank is recovered
once the $N-1$ paired comparisons 
$G_{N-1}=[(r_1,r_2), (r_2,r_3), ..., (r_{N-1},r_{N})]$
have been performed.
The probability to pick an element belonging
to $G_{N-1}$ is $2/N$, and from \req{t1} one needs doing so at least 
$N \log N$ times.
The minimal number of steps needed to find $\vec{r}=[r_1,r_2,...,r_N]$ 
with finite probability is then
\be\label{tc2}
t^c_{2} \simeq N^2 \log(c N),
\ee
where $c$ is a constant depending on $P$.

We should remark here that random comparisons are not at all
a good solution for sorting a vector of scalars: 
finding optimal sorting algorithms is a classic problem
of computer science \cite{knuth73}. In real situations, though,
one does not always have control over the evaluators' choice.
Besides, evaluators make mistakes and the outcome of a comparison
is often uncertain. In this case a common procedure is to perform 
round robins with successive eliminations, obtaining equation
\req{tc2} for the minimum number of comparisons needed to find
the winner \cite{LaMaZh04}.

{\bf Ranking a group of peers.}
Let us now consider a self evaluating community, a group
of $N$ users sharing the same expertise and voting for one
another.
%
%
This is equivalent to fixing $t=N$ in our general model.
The $N$ agents pick each $k$  randomly chosen peers, 
and establish a partial ranking among them 
according to their intrinsic values. 
We want to find, in the limit of large $N$, 
the minimal average value $k_r$ of $k$ that 
allows to recover the ``God-given'' ordering 
with  probability $P$.

\begin{figure}[h!]
\begin{center}
\includegraphics[width= \textwidth]{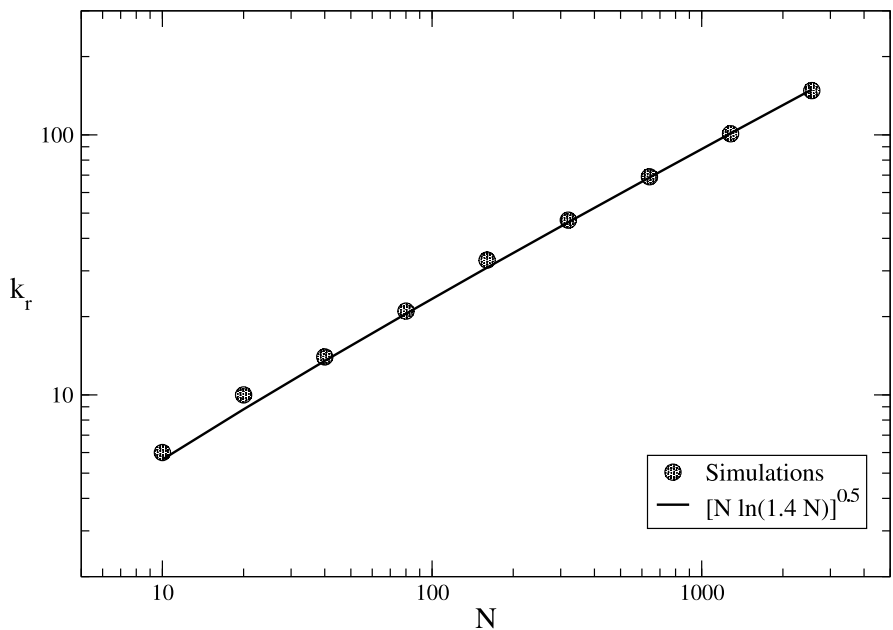}
\caption{Minimal average number of comparisons per agent $k_r$ 
needed to achieve the correct rank with probability $P=0.7$ as
a function of $N$. Circles are simulation results averaged over 100 realizations, the solid line
is the analytical estimation \req{kr}.}
\label{critikRel}
\end{center}
\end{figure}

When $k \ll N$
this model can be mapped into the previous one. In fact
making a local ordering of $k$ elements would require 
about $k^2$ paired comparisons. We then need $N k_r^2 \sim N^2 \log N$
comparisons to find the entire ranking. 
Thus
\bea\label{kr}
k_r  \simeq  \sqrt{N \log(c N)}. 
\eea
In figure \ref{critikRel} this last result is shown to match well
simulation data.

\section{Preferential attachment}
In real electronic communities fame plays an important role.
Objects that are already popular are more likely to receive comments,
books mentioned by important media are more likely to be
reviewed, people about whom many talk about are more likely 
to be judged again, and so on. Such ``richer get richer''
phenomenon, often referred to as preferential attachment \cite{AlBa02},
appears in numerous empirical data of social systems \cite{AlBaJe00} and is thought
to be one of the factors responsible for the emergence of scale free networks.
A direct consequence of this mechanism is that attendance is not
a fair indicator of web pages' quality \cite{BiLaYu04}.

Reputation systems collect evaluations about objects, aggregate and release them.
The aggregation process often consists in computing an average
of the received votes for each object. Assuming that objects
have an intrinsic quality $q_i$ and that
evaluations are random variables with mean equal to $q_i$,
the law of large numbers ensures that computed reputations
tend to the real intrinsic values once the number of users becomes large.
The implicit assumption is that every object receives a growing
number of evaluations, which is not always the case in the presence
of some sort of preferential attachment.

This can be verified within the framework of
the occupancy problem of section \ref{secRE}:
a user per time step $t$ evaluates $k=1$ objects among the $N$ available.
Instead of drawing the objects to be evaluated from a uniform
distribution, let us now define
\be\label{bb}
p_i(t)=\frac{v_i^{\alpha}(t)}{\sum\limits_{j=1}^N
v_j^{\alpha}(t)},
\ee
where $v_i(t)$ is the number of evaluations received by object $i$
up to time $t$, with $v_i(0)=1 \forall i$. The parameter $\alpha$ sets the
strength of the preferential attachment.
In the limit $\alpha \rightarrow \infty$ one has $p_i(t) \rightarrow \delta(i-i^*)$, 
where $i^* = \text{argmax}_i\{v_i(t)\})$; if $\alpha=0$, on the other hand,  $p_i$ becomes uniform.

As the real rank is not reached with certainty, it is useful to
define a distance on the rank space. Let
\be\label{distance}
d(r[1],r[2])=\frac{1-c(r[1],r[2])}{2}
\ee 
be the distance between two ranks, where 
$$
c(r[1],r[2])= \frac{\sum_{i=1}^N r_i[1] r_i[2] }{\sqrt{\sum_{i=1}^N r_i^2[1] \sum_{i=1}^N r_i^2[2]}} 
$$ 
is the rank correlation coefficient \cite{NumRec}.
\begin{figure}[ht!]
\begin{center}
\includegraphics[width= \textwidth]{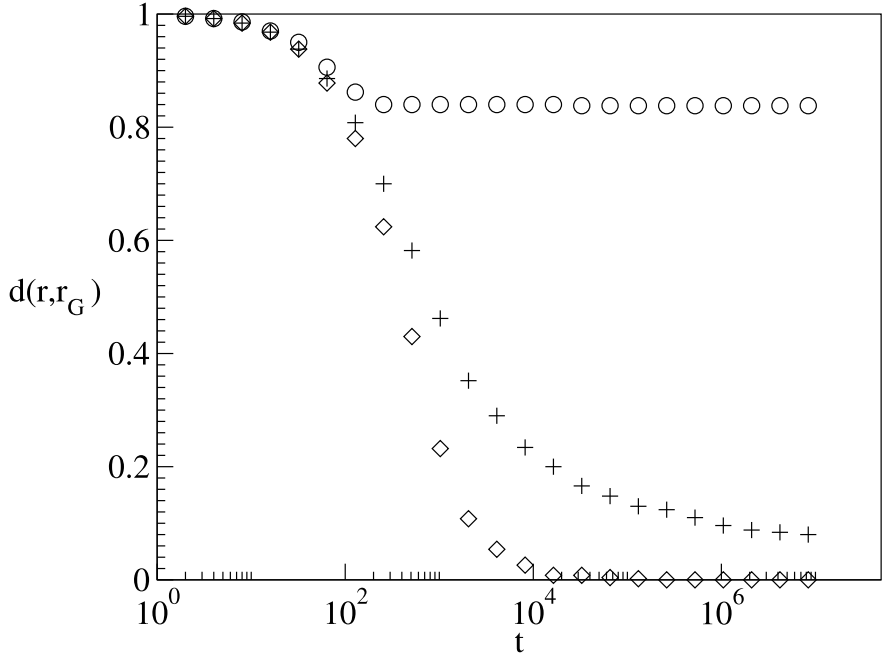}
\caption{
Time dependence of the distance \req{distance} between the ``God given rank'' $r_G$ and the rank $r$
  produced by the model with preferential attachment \req{bb}.
Symbols report, in log-linear scale, results of a simulation with
  $N=500$ agents and
$\alpha=2$ (circles), $\alpha=1$ (plus signs) and $\alpha=0$ (diamonds).
}
\label{QalphaAbs}
\end{center}
\end{figure}

In fig.\ref{QalphaAbs} the time dependence of the 
distance between the ``God given rank'' and the best guess $r$,
as defined in \req{distance}, is reported.
Numerical simulations show that for $\alpha>1$ the distance reaches a plateau $>0$, 
meaning that one can never achieve an arbitrary good estimate of qualities $\vec{q}$
because some agents never get to be evaluated. 
For $\alpha<1$, on the other hand,  the real rank is recovered
in a finite number of time steps; at the transition point
$\alpha=1$ one needs an infinite amount of steps
to evaluate all objects.

In fact we
can explain such a behavior with the following argument.
The transition appears when the number of
visits $v_i(t)$ does not grow in time for some $i$s.
For most objects the number of visits still grows linearly
in time
\be \label{vi}
v_i(t) \simeq f_i t
\ee
close to the transition.
We can thus approximate the denominator of \req{bb}
$\sum\limits_{j=1}^N v_j^{\alpha}(t) \simeq N t^{\alpha} I(\alpha)$,
where $I(\alpha)=\frac{1}{N}\sum\limits_{i=1}^{N} f^{\alpha}_i$.
The probability for a given object $i$ never to receive an
evaluation up to time $t$ is $Q_i(t)=\prod\limits_{t'=0}^{t}(1-p_i(t'))$,
the $\log$ of which, $\log(Q_i) \simeq -\frac{1}{NI(\alpha)}
\sum\limits_{s=1}^{t}\frac{1}{s^{\alpha}}$, only converges to a finite
value for $\alpha \ge 1$: the transition value is indeed $\alpha=1$.

\begin{figure}
\includegraphics[height=6cm, width=7cm]{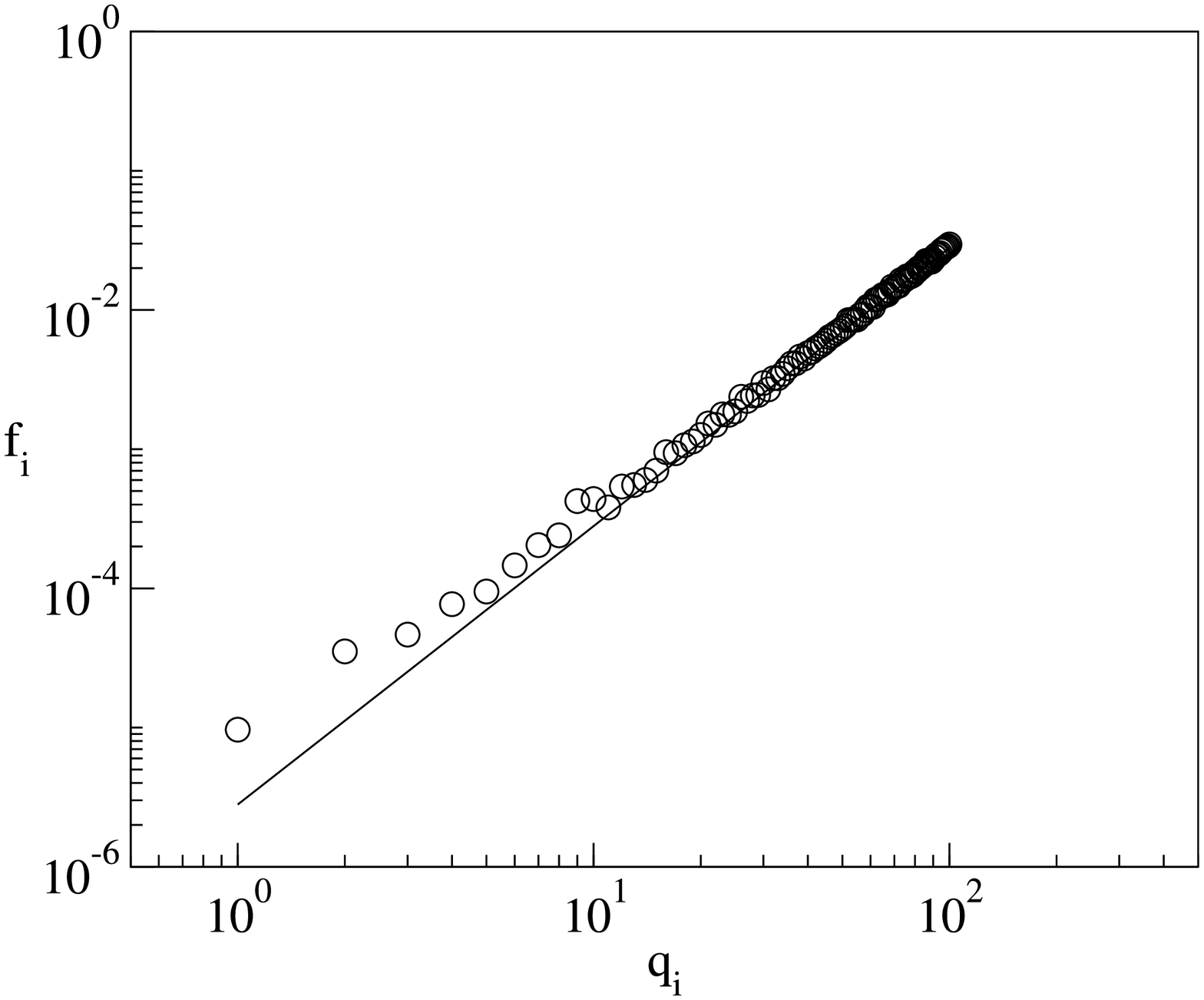}
\includegraphics[height=6cm, width=7cm]{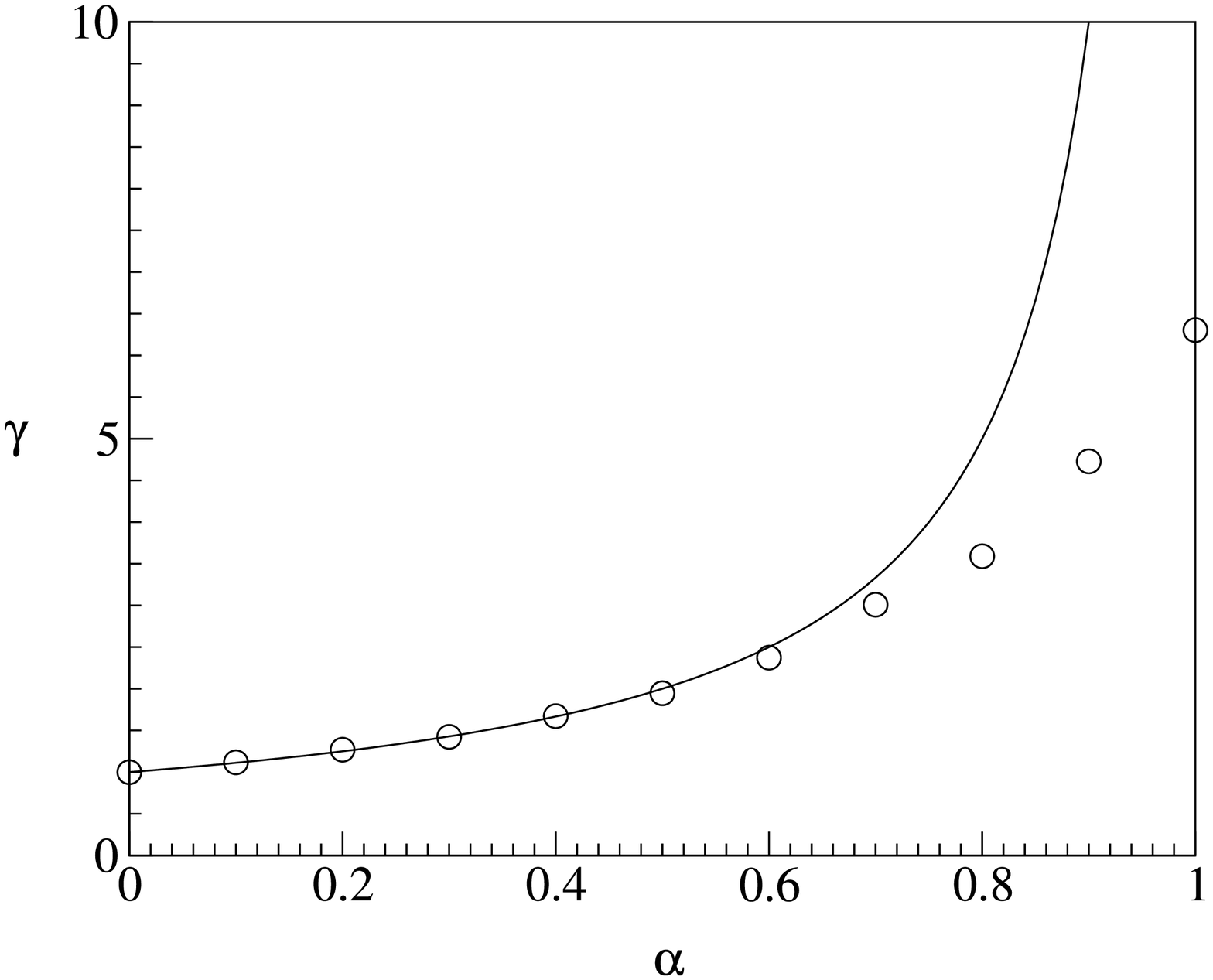}
\caption{(Left graph) Average rate of visits $f$ to an agent
plotted against her intrinsic quality $q$, after $10^5$ time steps, with $\alpha=0.5$.
(Right graph) Ratio $\gamma$ between the logarithm of the rate of visits $v_i/t$ 
and that of the intrinsic quality $q_i$
plotted against $\alpha$.
   Circles are simulation data,
   lines are calculated with equation \ref{feq}. 
Other simulatin parameters: 
number of agents $N=500$, number of realizations$=10$.
Qualities are uniformly distributed.}
\label{frelative}
\end{figure}

One can generalize the preferential attachment rule (\ref{bb}) 
by introducing multiplicative intrinsic qualities:
high quality agents have more chance to be evaluated, similarly to the model proposed
in ref. \cite{BiBa01}.
Let us define the probability to be evaluated as follows:
\be\label{bbq}
p_i(t)=\frac{q_{i}v_i^{\alpha}(t)}{\sum\limits_{j=1}^N
q_{j}v_j^{\alpha}(t)},
\ee
We can again approximate the denominator of \req{bbq} as $\sum\limits_{j=1}^N
q_jv_j^{\alpha}(t) \simeq N t^{\alpha} I(\alpha)$, where $I(\alpha)=\int\limits_{-\infty}^{+\infty}\rho(q) q
f^{\alpha}(q) dq$ and $\rho$ is the probability distribution of
the quality vector $q$. The same transition at $\alpha=1$ is found and
equation \req{vi} yields 
\bea\label{feq}
f_i &=& \left(\frac{q_{i}}{NI(\alpha)}\right)^{\gamma}\\ \label{gamma}
\gamma &=& \frac{1}{1-\alpha}.
\eea
Numerical simulations are shown to agree with  this approximation
in figure \ref{frelative}. 
The left panel displays the $q$ dependence of $f$ for fixed $\alpha$:
equation (\ref{feq}) asymptotically matches the simulation data. 
In the right panel of figure \ref{frelative} the
ratio $\gamma$ between the logarithm of the rate of visits $v_i/t$ 
and that of the intrinsic quality $q_i$
is plotted against $\alpha$. 
Equation \req{gamma} fits the data for $\alpha <\alpha_c$
and diverges at $\alpha=1$ as expected. The simulations
were carried on with a uniform $\rho(q)$.

\begin{figure}[ht!]
\begin{center}
\includegraphics[width= \textwidth]{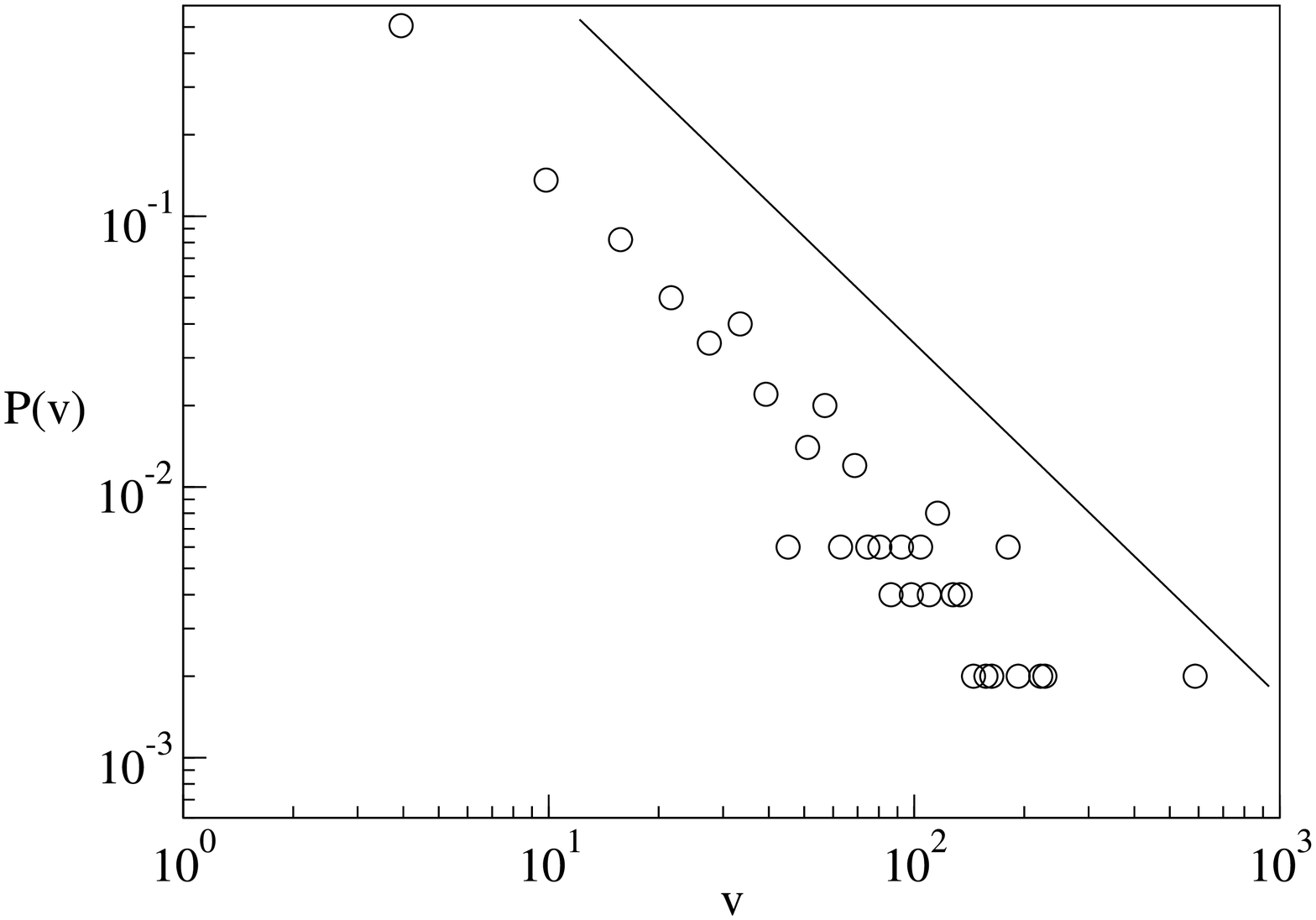}
\caption{Probability distribution $P(v)$ of the number of visits.
Circles are simulation data of the model characterized by equation
\req{bbq} with   $\alpha = 1$ and $N=500$, after $10^5$ steps.  
Here qualities are uniformly spaced, i.e. $q_i=i$.
The solid line is a power law with exponent $-1.3$.}
\label{visit}
\end{center}
\end{figure}

The distribution of the number of visits 
becomes a power law around the transition point $\alpha=1$, as
reported in figure \ref{visit}. 

The results described in this section can be easily extended to the 
case of pairwise comparisons introduced in section
\ref{secRE}. That is, agents compare $2$ objects per time step,
chosen according to equation  \req{bbq}, and
rank them in order of decreasing quality. The global ranking
is finally established on the basis of these partial orderings.
The question is to find the maximal value of $\alpha$ that allows
to recover the true rank with finite probability.
This can be done by evaluating the probability that two agents 
(belonging to the set $G_{N-1}$)
will never be compared. By very similar reasoning to the
absolute judgment case, one obtains $\alpha_c = \frac{1}{2}$. 

\section{Aknowledgments}
For early collaboration on this paper
and sharing insights we thank Hassan Masum and Yi-Kuo Yu; for discussions
and a careful reading of the manuscript we thank Frantisek Slanina and
Giancarlo Mosetti. This work was partially supported by the Swiss
National Science Foundation through project number 2051-67733.

\bibliographystyle{elsart-num}
\bibliography{vot}

\end{document}